\documentclass[12pt,preprint]{aastex}
\usepackage{graphicx}
\usepackage{times}
\usepackage{float}
\usepackage{wrapfig}
\usepackage{float}
\usepackage{rotating}
\usepackage{lscape}
\usepackage{longtable}
\usepackage{CJK}

\shorttitle{Debris Disk Variability}
\shortauthors{Meng et al. (2012)}

\begin{document}
\begin{CJK*}{UTF8}{gbsn}

\title{Variability of the Infrared Excess of Extreme Debris Disks}

\author{Huan Y. A. Meng (孟奂)\altaffilmark{1}, George H. Rieke\altaffilmark{1,2}, Kate Y. L. Su\altaffilmark{2}, Valentin D. Ivanov\altaffilmark{3}, Leonardo Vanzi\altaffilmark{4}, Wiphu Rujopakarn\altaffilmark{2}}

\altaffiltext{1}{Department of Planetary Sciences, Lunar and Planetary Laboratory, University of Arizona, Tucson, AZ 85721, USA}
\altaffiltext{2}{Steward Observatory, University of Arizona, Tucson, AZ 85721, USA}
\altaffiltext{3}{European Southern Observatory, Ave. Alonso de C\`{o}rdova 3107, Casilla 19, Santiago, 19001, Chile}
\altaffiltext{4}{Department of Electrical Engineering and Center of Astro Engineering, Pontificia Universidad Catolica de Chile, Av. Vicu\~{n}a Mackenna 4860, Santiago, Chile}

\begin{abstract}
Debris disks with extremely large infrared excesses (fractional luminosities $> 10^{-2}$) are rare. Those with ages between 30 and 130 Myr are of interest because their evolution has progressed well beyond that of protoplanetary disks (which dissipate with a timescale of order 3 Myr), yet they represent a period when dynamical models suggest that terrestrial planet building may still be progressing through large, violent collisions that could yield large amounts of debris and large infrared excesses. For example, our Moon was formed through a violent collision of two large proto-planets during this age range. We report two disks around the solar-like stars \objectname{ID8} and \objectname{HD 23514} in this age range where the 24 $\micron$ infrared excesses vary on timescales of a few years, even though the stars are not variable in the optical. Variations this rapid are difficult to understand if the debris is produced by collisional cascades, as it is for most debris disks. It is possible that the debris in these two systems arises in part from condensates from silicate-rich vapor produced in a series of violent collisions among relatively large bodies. If their evolution is rapid, the rate of detection of extreme excesses would indicate that major collisions may be relatively common in this age range.
\end{abstract}

\keywords{circumstellar matter --- infrared: stars --- stars: individual (2MASS J08090250-4858172, HD 23514) --- planets and satellites: formation}

\section{Introduction}

Circumstellar disks are key signposts for planetary system evolution both because they are associated closely with planets, and because they are much more readily observed than the planets themselves. Optically thick protoplanetary disks are born around young stars and dissipate over a time $< 10$ Myr \citep[e.g.,][]{wya08}. Afterwards, optically thin planetary debris disks can emerge, sustained by the fragmentation of colliding planetesimals \citep{wya08}. Debris disks are readily detected through their infrared excess emission (over the stellar photospheric output), which is produced by dust warmed by the star. Such excesses are found over the entire age range of main sequence stars to ages of $\sim10$ Gyr. Therefore, they can be used to probe the presence and state of planetesimal systems (such as the analogs of asteroid and Kuiper Belts) and, indirectly, the evolution of the accompanying planetary systems. In particular, debris disks can be used to search for phases in the evolution of other planetary systems that played major roles in the evolution of the Solar System and in the formation of terrestrial planets.

One such phase extends roughly from 30 to 130 Myr, well separated from the era of protoplanetary disks but when dynamical models \citep[e.g.,][]{cha01,mor10} predict that giant impacts will still occur \citep{rig11}. For example, our Moon formed during this period through a massive collision between the proto-Earth and a Mars-sized proto-planet \citep{can04}. Such events should yield huge amounts of debris; indeed, a few candidates have been found. They take the form of infrared excesses from heated debris that far exceed those expected from the trend for quiescently evolving systems \citep[see summary in][]{bal09}. These extreme systems (defined as total emission $>$ four times the stellar photospheric output at 24 $\micron$) occur around only $\sim1{\%}$ of solar-like (F5-G9) stars \citep{bal09}. If evolutionary timescales similar to those of conventional debris systems are adopted \citep[on the order of 3 Myr,][]{gro01}, then the low level of incidence of such excesses implies that major collisions are rare.

In this paper, we present {\it Spitzer} data that show that two of these extreme systems, \objectname{ID8} in \objectname{NGC 2547} and \objectname{HD 23514} in the \objectname{Pleiades} (both with fractional debris disk luminosities $\gtrsim 2 \times 10^{-2}$), vary significantly at 24 $\micron$ over a few years. The age of NGC 2547 is $\sim$35 Myr \citep{jef05}, while that of the Pleiades is $\sim$120 Myr \citep{sta98}. Despite their large and variable infrared excesses, both \objectname{ID8} and \objectname{HD 23514} are well past the age of protoplanetary disks and the variations we report are unlikely to be related to those found to be ubiquitous in the latter disk type at ages $< 10$ Myr \citep[e.g.,][]{esp11,fla12}. \citet{mel12} have reported a third example of strong variability in an extreme infrared excess in the same 30 $-$ 130 Myr age range, so the phenomenon appears to be characteristic of this particular class of circumstellar disk.

\objectname{ID8} is a type G6-G7 dwarf (N. Gorlova et al., in preparation) at a distance of $\sim$450 pc and with optical colors indicating negligible extinction. \objectname{HD 23514} is of F5V spectral type \citep{gra01} at a distance of $\sim$130 pc, with normal colors for this type \citep{men67} when corrected for the foreground reddening \citep{cer87}. \objectname{ID8} is possibly a binary star based on a radial velocity measurement 8 km s$^{-1}$ different from the cluster mean (N. Gorlova et al., in preparation); a late-M companion to \objectname{HD 23514} has recently been discovered at a projected distance of 360 AU from the star \citep{rod12}. Neither star is variable in the optical, to within the $\sim1\%$ measurement errors.

\section{Data}

{\it Spitzer}/MIPS observations of \objectname{ID8} at 24 $\micron$ were obtained in programs for G. Rieke (PID 58) and C. Lada (PID 20124), The data for \objectname{HD 23514} are from programs led by M. Meyer (PID 148) and G. Rieke (PIDs 30503, 30566). We also show in Figure 1 IRS spectra; that for \objectname{ID8} is from PID 40227, PI G. Rieke, while that for \objectname{HD 23514} is from PID 50228, PI I. Song. In summary, we have three epochs of MIPS 24 $\micron$ observations for both targets, supplemented with a 4th epoch observation using IRS spectra. In addition, one epoch of IRAC observation is available for each target, from PID 58 for \objectname{ID8} (PI G. Rieke) and PID 50228 for \objectname{HD 23514} (PI I. Song).

The 24 $\micron$ data were reduced with the MIPS instrument team Data Analysis Tool \citep{gor05}. In addition, a second flat field constructed from the 24 $\micron$ data itself was applied to remove scattered-light gradients and dark latency \citep{eng07}. We extracted the photometry using PSF (point spread function) fitting. The input PSFs were constructed using observed calibration stars and smoothed STinyTim model PSFs \footnote{Krist, J., 2006, Tiny Tim / Spitzer User's Guide, Version 2.0}, and have been tested to insure the photometry results are consistent with the MIPS calibration \citep{eng07}. The final photometry errors also include the errors from the detector repeatability \citep[$\leq$ 1{\%} at 24 $\micron$,][]{eng07}. These errors should apply directly to \objectname{HD 23514}, which is similar in brightness to the standard stars described in \citet{eng07}. We measured an isolated, similar brightness star $\sim$4$\arcmin$ south of \objectname{ID8} to test the 24 $\micron$ photometry stability among the three epochs of data. The measured 24 $\micron$ flux (5.46 mJy) is constant within 1.5\% (rms), consistent with the repeatability of MIPS photometry on brighter stars. The measured flux densities and 24 $\micron$ magnitudes ([24]) assume 7.17 Jy as the zero magnitude flux density.

The IRAC photometry of \objectname{ID8} has been published by \citet{gorl07}. For \objectname{HD 23514}, the IRAC BCD images were retrieved from the {\it Spitzer} archive, and mosaicked with the MOPEX software with a scale of 0.6\arcsec\ pixel$^{-1}$. The final mosaic images were checked against the archival PBCD images for consistency. Aperture photometry was then performed on the verified mosaic images with an aperture radius of 15\arcsec\ and sky annulus between 24\arcsec\ and 36\arcsec\ from the opto-center. The flux conversion issue in the S18.18 IRAC processing\footnote{http://irsa.ipac.caltech.edu/data/SPITZER/docs/irac/s18.18fluxconv.shtml} is corrected.

IRS spectra were reduced and extracted using the SMART software \citep{hig04,leb10}. We also computed synthesized 8.0 and 24 $\micron$ photometry by integrating the IRS spectra and comparing with similar integrals over an A-star spectrum. We adopt 5\% uncertainty for the synthesized photometry which is consistent with the cross calibration among IRAC, MIPS, and IRS \citep{cush06,car10}. The 24 $\micron$ photometry is summarized in Table 1 and shown in Figure 1. Both stars are found to vary at 24 $\micron$ at a high degree of significance.

Visible photometry was collected from the literature (Table 2). Neither star has been noted to vary; in fact, \objectname{HD 23514} has been used as a local photometric standard to study variations in neighboring stars \citep{alp81}. Because there were only four relevant measurements for \objectname{ID8}, we obtained two more; eight measurements were already available for \objectname{HD 23514}. The data for both stars show a scatter of only $\pm 1\%$, consistent with measurement errors and corroborating the lack of variability at visible wavelengths.

WISE observations of \objectname{ID8} were performed in mid May, 2010, nearly 6.5 years after the IRAC observation. Here we consider only the W1 and W2 bands because of their spectral proximity to the two short wavelength bands in IRAC. We randomly select three other \objectname{NGC 2547} member stars whose $JHK$, 3.6 and 4.5 $\micron$ brightnesses are all similar to that of \objectname{ID8}. The ($K_s$ - $W2$) colors of the three comparison stars are in a narrow interval between 0.00 and 0.03, while that of \objectname{ID8} is approximately 0.67, confirming the 4.5 $\micron$ excess. On the other hand, the ([4.5] - $W2$) colors of the comparison stars range from 0.01 to 0.05, while that of \objectname{ID8} is 0.28. Given the similarity of the IRAC 4.5 $\micron$ and WISE $W2$ bands, the significant difference between the fluxes in these bands at different epochs is most easily explained as the result of variability. The same trend may also be observed at 3.6 $\micron$ (or $W1$) at a low degree of significance. However, variations in the 3 - 4 $\micron$ region are not seen for \objectname{HD 23514} over the $\sim$1.5 year time span between the IRAC and WISE observations. 

In summary, we found 10\% $-$ 30\% peak-to-peak variation at 24 $\micron$ on yearly timescales while no changes were found at optical wavelengths for these two systems. The change between IRAC and WISE, as well as the IRS synthetic photometry, suggests that variations are also possibly seen at 4.5 and 8.0 $\micron$.

\section{Discussion}

\objectname{HD 23514} and a few other of extreme-excess stars \citep[e.g., \objectname{HD 15407},][]{fuj09,mel10} show spectra in the 10 $\micron$ region (see Figure 1) indicative of the presence of finely divided silica dust \citep{lis09} or analogs of silica ``smoke'' \citep{kim07}. The spectrum of \objectname{ID8} is also complex, with features indicating significant amounts of finely divided crystalline silicates (N. Gorlova et al., in preparation). The large equivalent widths of the features in these spectra require grain sizes of order 0.1 $\micron$ \citep{bou08}, well below the blowout size for these stars. The dust properties suggest that the infrared excesses of these stars are associated with violent, recent events \citep{lis09} and hence may differ significantly from typical planetary debris systems whose mid-infrared spectra tend to be featureless \citep{wya08} and whose evolution is dominated by an overall slow monotonic decay with time \citep{rie05,wya08}. \objectname{BD +20 307} \citep[][several Gyr old]{son05}, which also has a very strong emission band around 10 $\micron$, may also vary by $\sim$7\% in the WISE $W3$ and $W4$ bands (12 and 22 $\micron$) between two scans $\sim$188 days apart, although it is probably stable at $W1$, suggesting that the mid-infrared variability might be correlated with the presence very fine dust particles.

Traditional planetary debris disks are sustained by collisional cascades in which populations of planetesimals collide until they are ground down to micron-sized dust, when they are ejected by radiation pressure or spiral into the star due to Poynting-Robertson drag. The timescale for the overall changes in the infrared excesses in these systems is expected to be millions of years \citep{wya08}. However, around stars of solar mass or less, theoretical models by \citet{ken05} indicate that short-term spikes in the 24 $\micron$ output can occur due to large collisions (of $\sim$100 km bodies). Thus, the variations support the arguments that we are seeing the consequences of individual collisional events and the resulting rapid generation of large swarms of particles.


We can estimate the timescales for such events from the scaling relations given by \citet{wya08}. To apply these relations, we need to estimate the location of the planetesimal belts around the two stars. Because small grains associated with recent transient events dominate the mid-infrared properties, it is not possible to derive the properties of the parent body planetesimal system from the mid-IR measurements. Instead, we assume that the parent body populations are similar to those around stars of similar spectral type and age. The infrared excesses at 5.8 and 8.0 $\micron$ trace dust within $\sim$1 AU of a solar-type star. We have complemented studies of these excesses by \citet{gorl07, car09} by analyzing the data for the \objectname{Pleiades} from \citet{sta07}. In the latter case, we found that none of the stars with [3.6] $\le$ 10 (corresponding to solar-like absolute magnitudes) have excesses. The net result is that, in the age range (30 - 130 Myr), there are only two cases other than \objectname{ID8} and \objectname{HD 23514} with reliably detected excesses at 5.8 $\micron$ and 8.0 $\micron$ out of a total of 352 solar-like stars observed. This result shows that planetesimal belts in debris systems in this age range are nearly always located outside 1 AU. If into equations 14 - 16 of \citet{wya08}, we substitute solar masses and luminosities, a fractional luminosity of $2 \times 10^{-2}$, a radius of 1 AU, $dr / r$ of 0.5, and eccentricity of 0.2 (values to yield as rapid evolution as possible), we find that the time to remove a 100 km diameter body by collisional cascade is $t_c \gtrsim 100$ years. On the other hand, the time to remove 1 mm particles is $t_c \sim 0.1$ year, an order of magnitude shorter than the time span of the observations. This suggests that small particles are replenished by secondary collisions. Given that the variations are $\sim$10\% of the fractional luminosity, the mass associated with them is $\sim$10$^{-2}$ M$_{\earth}$, integrating the particle size distribution from 0.1 $\mu$m to 100 km. Significantly lower mass estimates are only possible if the planetesimal belt is well inside 1 AU.

\citet[][Figure 19]{gro01} present detailed models that find a characteristic timescale of $3 \times 10^6$ years for the decay of dust from collisions in the asteroid belt. The behavior is supported by the role of recent collisions in the production of zodiacal dust bands \citep{nes03}. This result provides an independent test of the timescale for decay of excess due to collisional cascades around \objectname{ID8} and \objectname{HD 23514}. For the population of disks around stars of similar mass and age to these two, a typical fractional luminosity in a quiescent state is $5 \times 10^{-4}$ \citep{car09}. Most of this emission is from cool dust that must be well outside a few AU from the star. The fractional luminosity for warm dust is an order of magnitude lower \citep{car09}, i.e. $5 \times 10^{-5}$. These values, the scaling relations in \citet{wya08}, a solar fractional luminosity of $10^{-7}$ \citep{bac93}, and orbital radius of 2.5 AU for the zodiacal cloud and asteroid belt indicate $\sim$100 years as the characteristic time for decay of a transient event via collisional cascade at 1 AU around \objectname{ID8} or \objectname{HD 23514}. This value agrees well with that obtained directly from the scaling relations.

Although this estimate is rough, it indicates that generating the variable excesses through collisional cascades may not be impossible, but would require extreme assumptions. We therefore will explore alternatives. The possibility that the 24 $\micron$ changes reflect varying rates of heating by the star is made unlikely by the complete lack of variations in the visible for both stars. Phenomena that avoid any kind of collisional cascade will inherently have faster timescales. One such process might be a collision that caused a large body to lose its regolith, injecting pre-processed small dust into the interplanetary space around the star. However, very small grains tend to be absent in planetesimal regoliths because radiation pressure slowly drives the small particles away \citep{mas09}; the regolith hypothesis is possibly contradicted by the evidence for small grains in the mid-IR spectra. In addition, regoliths have a full size spectrum of objects \citep{hou82}, so such a regolith ejection event would be analogous to launching a fully developed collisional cascade and it would have a similar evolutionary timescale to the cascades already considered. An alternative might be the disintegration of one or several extremely large comets. The most finely divided dust released would have a short dwell time near the star, since it would be ejected by radiation pressure or lost through Poynting-Robertson drag; the larger bodies would remain for some time because the comet orbit would not coincide with a dense planetesimal belt and would have inadequate density of objects for vigorous collisional activity. The timescale for the first process is probably too short, making the resulting excesses too rare to agree with the simultaneous detection of a number of them, and that for the second too long to explain the variability of \objectname{ID8} and \objectname{HD 23514}.

None of these possibilities can convincingly be shown to produce the huge amounts of dust around these stars while having the necessary short timescales for variability. Therefore, we consider another possibility -- that the dust condenses from silica-rich vapor associated with violent collisions among large bodies. For example, in massive collisions such as the one that formed our Moon, of order 20{\%} of the mass emerges as vapor rich in silica \citep{can04}. The production of silica-rich vapor can occur at collisional velocities above about 2 km s$^{-1}$ based on simple energetic arguments. However, a threshold of about 10 km s$^{-1}$ is generally adopted because a substantial portion of the energy is deposited in collisionally produced fragments \citep{hor00}. Laboratory experiments suggest that the vapor will condense quickly into silica ``smoke'' \citep{kim07}, and forsterite-like crystals are the most common crystalline form to grow from the condensation products \citep{kob08}. The spectrum of silica smoke \citep{kim07} resembles that of the infrared excess of \objectname{HD 23514} \citep{rhe08}, while the small grains that dominate the spectrum of \objectname{ID8} are rich in forsterite-like materials (N. Gorlova et al., in preparation). The resulting small particles will be cleared quickly by radiation pressure and the Poynting-Robertson effect. The cooling and disk spreading timescale for the silica-rich protolunar disk that later formed the Moon is $\sim$one year \citep{tho88,can04}. However, most of the ejected mass in the moon-forming impact remained gravitationally bound to the Earth; the small solid angle viewed from the Sun indicates that such a single event could not raise the infrared luminosity high enough to explain the excesses in these systems. However, the significant amount of escaping material from a major impact might initiate a violent collisional cascade and elevate the infrared excess of the star for a longer period. Thus, the extreme excesses around ID8 and HD 23514 are unlikely to be due to an individual violent collision, but to reflect extended episodes of such collisions possibly triggered by a single event.


\section{Conclusion}

Dynamical models indicate that giant impacts continue to build terrestrial planets into the 30 - 130 Myr age range. A small number of stars have extreme infrared excesses and mid-infrared spectra that suggest that such impacts have occurred recently around them. However, the duration of these indicators, and hence the frequency of events required to account for the numbers we detect, is not clear. We report that the excess is variable on year-timescales around two of these stars, \objectname{ID8} in \objectname{NGC 2547} and \objectname{HD 23514} in the \objectname{Pleiades}; a third variable example has been reported by \citet{mel12}. The variability is too rapid to be explained readily if the excesses are generated in collisional cascades, as is the case for most planetary debris disks. Instead, it is possible that the dust producing the excess condenses from vapor generated in violent collisions. The timescale for variations of years, rather than millions of years, suggests that the evolution of the infrared excesses might be rapid. In that case, major collisions would need to occur frequently to account for the number of stars in this age range that are observed to have extreme excesses.
 
\acknowledgments

This work is based on observations made with the Spitzer Space Telescope, which is operated by the Jet Propulsion Laboratory, California Institute of Technology, under NASA contract 1407. Support for this work was provided by NASA through contract 1255094, issued by JPL/Caltech. Optical observations obtained partly at Observatory UC Santa Martina, and with facilities supported by the National Astronomical Research Institute of Thailand (NARIT).

{\it Facilities:} \facility{Spitzer (IRAC, IRS, MIPS), WISE}

\clearpage

\begin{deluxetable}{ccccccc}
\tabletypesize{\scriptsize}
\tablewidth{15cm}
\tablecaption{Infrared Photometry by {\it Spitzer} and WISE \label{IR}}
\tablecolumns{7}
\tablehead{
\colhead{Date} & \colhead{3.6 $\micron$ (mJy)} & \colhead{4.5 $\micron$ (mJy)} & \colhead{5.8 $\micron$ (mJy)} & \colhead{8.0 $\micron$ (mJy)} & \colhead{24 $\micron$ (mJy)} & \colhead{source}}
\startdata
\multicolumn{7}{c}{\objectname{ID8}} \\
\hline
2003 Dec 4  & 9.74 $\pm$ 0.1  & 7.35 $\pm$ 0.1  & 6.56 $\pm$ 0.1                   & 7.64 $\pm$ 0.1         & \nodata               & \citet{gorl07}\\
2004 Jan 29 & \nodata         & \nodata         & \nodata                          & \nodata                & 6.23 $\pm$ 0.10       & AOR 4318976\\
2006 May 9  & \nodata         & \nodata         & \nodata                          & \nodata                & 8.34 $\pm$ 0.12       & AOR 16798464 \\
2006 Dec 4  & \nodata         & \nodata         & \nodata                          & \nodata                & 7.32 $\pm$ 0.11       & AOR 16800512\\
2007 Jun 16 & \nodata         & \nodata         & \nodata                          & 6.38\tablenotemark{a}  & 6.74\tablenotemark{a} & AOR 21755136 \\
2010 May 13-21 & 12.05 $\pm$ 0.26\tablenotemark{b} & 9.11 $\pm$ 0.17\tablenotemark{b} & \nodata & \nodata & \nodata & WISE All-Sky Catalog \\
\cutinhead{\objectname{HD 23514}}
2004 Sep 22 &\nodata          & \nodata         & \nodata                          & \nodata                & 67.7 $\pm$ 0.7        & AOR 5320192\\
2007 Feb 25 & \nodata         & \nodata         & \nodata                          & \nodata                & 68.0 $\pm$ 0.7        & AOR 18303232 \\
2007 Sep 17 & \nodata         & \nodata         & \nodata                          & \nodata                & 61.6 $\pm$ 0.7        & AOR 17656576\\
2008 Sep 18 & 206.3 $\pm$ 1.1 & 180.7 $\pm$ 1.2 & 154.6 $\pm$ 0.7                  & 203.9 $\pm$ 1.0        & \nodata               & AOR 25790208\\
2008 Oct 1  & \nodata         & \nodata         & \nodata                          & 169.3\tablenotemark{a} & 64.8\tablenotemark{a} & AOR 25789952\\
2010 Feb 12-13 & 225.3 $\pm$ 4.5\tablenotemark{b} & 180.6 $\pm$ 3.1\tablenotemark{b} & \nodata & \nodata & \nodata & WISE All-Sky Catalog \\
\enddata
\tablenotetext{a}{Synthetic photometry based on IRS spectrum and the response curves of IRAC and MIPS. The cross-calibration uncertainties of the {\it Spitzer} instruments are expected to be $\sim5\%$ \citep{car10}.}
\tablenotetext{b}{Flux densities at WISE $W1$ and $W2$. The effective wavelengths are 3.35 and 4.60 $\micron$, slightly different from those of IRAC.}
\end{deluxetable}

\clearpage

\begin{deluxetable}{ccc}
\tabletypesize{\scriptsize}
\tablewidth{5cm}
\tablecaption{Visible Photometry\label{vis}}
\tablecolumns{3}
\tablehead{
\colhead{V} & \colhead{I\tablenotemark{a}} & \colhead{references}
}
\startdata
\multicolumn{3}{c}{\objectname{ID8}} \\
\hline
13.14 & 12.34 & 1 \\
\nodata & 12.34 & 2 \\
13.12 & 12.34 & 3 \\
\nodata & 12.32 & 4 \\
13.12 & \nodata & 5 \\
13.14 & \nodata & 5 \\
\cutinhead{\objectname{HD 23514}}
9.42 & \nodata & 6 \\
9.42 & 8.67 & 7 \\
9.44 & \nodata & 8 \\
9.43 & \nodata & 9 \\
9.42 & \nodata & 10 \\
9.44 & \nodata & 11 \\
9.43 & \nodata & 12 \\
9.44 & \nodata & 13 \\
\enddata
\tablenotetext{a}{Cousins filter for \objectname{ID8} and Johnson filter for \objectname{HD 23514}.}
\tablerefs{
(1) Naylor et al. 2002; (2) Jeffries et al. 2004; (3) Lyra et al. 2006; (4) DENIS catalog;
(5) this work, observed on 2011 Nov 18;
(6) Johnson \& Mitchell 1958; (7) Mendoza 1967; (8) Rufener 1976;
(9) Alphenaar \& van Leeuwen 1981; (10) Cernis 1987; (11) van Leeuwen et al. 1986;
(12) Tycho catalog, transformed from \citet{bes00}; (13) Droege 2006.
}
\end{deluxetable}

\clearpage

\begin{figure}
\epsscale{1}
\label{fig1}
\plotone{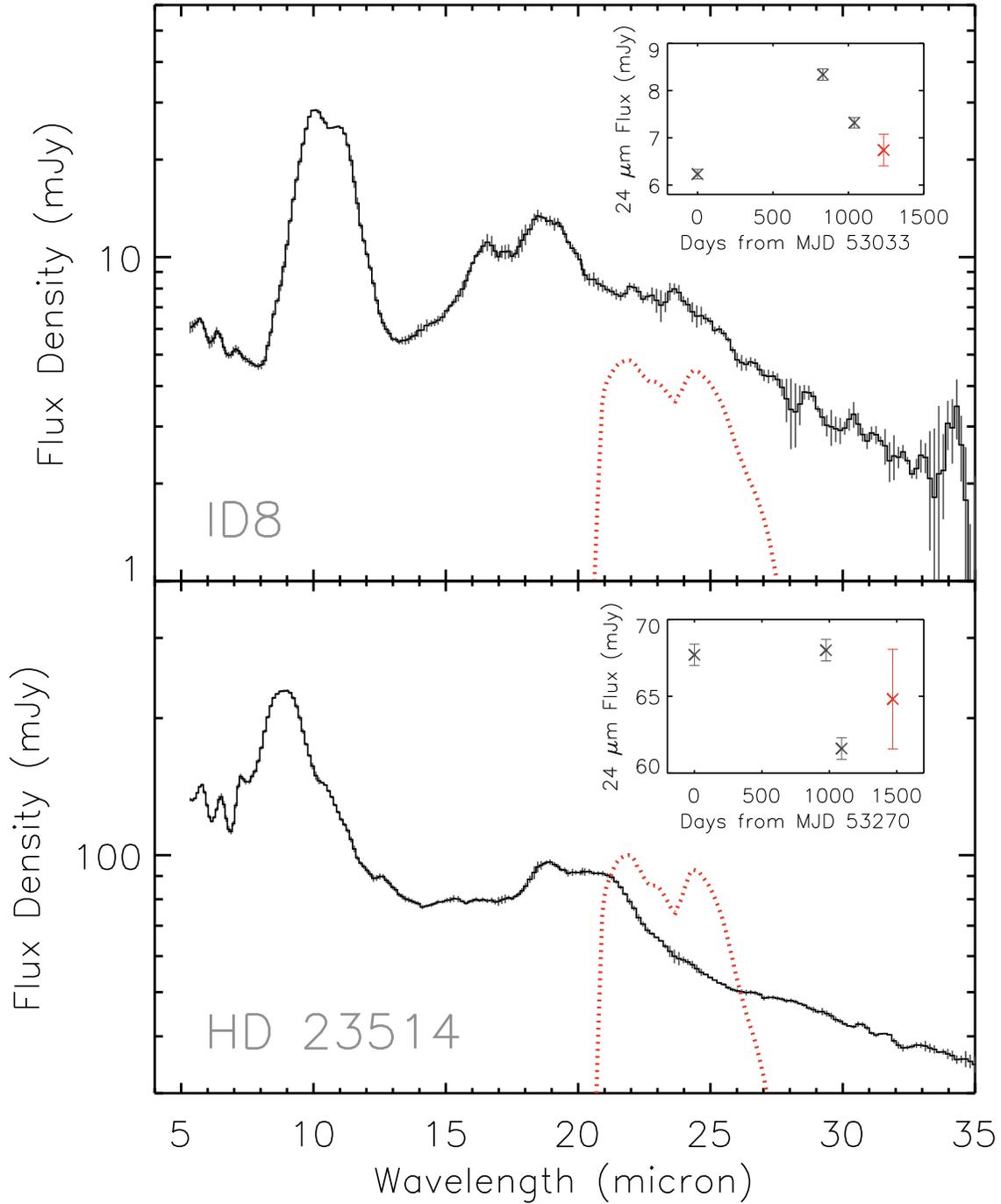}
\caption{Top: IRS low-resolution spectrum of ID8 (smoothed by three pixels), with our MIPS photometric points at 24 $\micron$ (black) plus a value from synthetic photometry of the spectrum (red). A 5\% uncertainty is adopted for the synthetic photometry. The response curve of MIPS 24 $\micron$ is overplotted for comparison. Bottom: Same, for HD 23514.}
\end{figure}

\end{CJK*}
\end{document}